\documentclass[preprint,12pt]{elsarticle}

\usepackage{color}

\usepackage{amssymb}
\usepackage{hyperref}

\journal{JQSRT}

\begin{document}

\begin{frontmatter}

\title{Cross-sections for heavy atmospheres: H$_2$O self-broadening}

\author[inst1]{Lara O. Anisman}
\author[inst2]{Katy L. Chubb}
\author[inst1]{Quentin Changeat}
\author[inst3,inst1]{Billy Edwards}
\author[inst1]{Sergei N. Yurchenko}
\author[inst1]{Jonathan Tennyson}
\author[inst1]{Giovanna Tinetti}

\affiliation[inst1]{organization={Department of Physics and Astronomy, University College London},
            addressline={Gower Street}, 
            city={London},
            postcode = {WC1E 6BT},
            country={United Kingdom}}

\affiliation[inst2]{organization={Centre for Exoplanet Science, University of St Andrews},
            addressline={North Haugh}, 
            city={St Andrews},
            postcode={KY16 9SS},
            country={United Kingdom}}
            
\affiliation[inst3]{organization={AIM, CEA, CNRS, Universit\'e Paris-Saclay, Universit\'e de Paris},
            city={Gif-sur-Yvette},
            postcode={F-91191},
            country={France}}
         
\begin{abstract}

The discovery of super-Earth and mini-Neptune exoplanets means that atmospheric signals from low-mass, temperate exoplanets are being increasingly studied. The signal acquired as the planet transits its host star, known as the \textit{transit depth}, is smaller for these planets and, as such, more difficult to analyze. The launch of the space telescopes \emph{James Webb} (JWST) \& \emph{Ariel}  will give rise to an explosion in the quality and quantity of spectroscopic data available for an unprecedented number of exoplanets in our galaxy. Accurately extracting the information content, thereby permitting atmospheric science, of such data-sets will require robust models and techniques. We present here the analysis of simulated transmission spectra for water-rich atmospheres, giving evidence for non-negligible differences in simulated transit depths when self-broadening of H$_2$O is correctly accounted for, compared with the currently typically accepted standard of using H$_2$ and He-broadened cross-sections. Our case-study analysis is carried out on two super-Earths, focusing on water-based atmospheres, ranging from H$_2$-rich to H$_2$O-rich.  The transit depth is considerably affected, increasing values by up to 60 ppm, which is shown to be detectable with JWST and Ariel. The differences are most pronounced for the lighter (i.e. $\mu \sim 4$)  atmospheres. Our work illustrates that it is imperative that the field of exoplanet spectroscopy moves toward adapted cross-sections, increasingly optimized for high-$\mu$ atmospheres for studies of super-Earths and mini-Neptunes.
\end{abstract}

\begin{keyword}
exoplanet atmospheres \sep water vapor \sep opacities \sep radiative transfer \sep line broadening \sep super-Earths \sep mini-Neptunes

\end{keyword}

\end{frontmatter}

\section{Introduction}
The field of exoplanet atmospheric spectroscopy relies heavily upon accurately derived cross-sections, generated for particular pressure and temperature ranges as well as for specific molecules.
As the number of known exoplanets continues to increase rapidly, we are witnessing an influx of small, more temperate worlds ($\leq 10 $ $M_{\bigoplus}, \leq 5000$ K).
The range of super-Earths and mini-Neptunes discovered has opened up the field of possibilities for observable atmospheres on these planets. Prime examples include the TRAPPIST-1 system, \citep{de_Wit_2018}, 55 Cnc\,e \citep{tsiaras_55cnce, Demory_2016}, GJ\,1132\,b \citep{libby_gj1132,aresV_gj1132}, GJ 1214\,b \citep{Kreidberg_GJ1214b_clouds}, K2-18\,b \citep{Tsiaras_2019_k2-18} and LHS 1140\,b \citep{Edwards_2020}.

Small planets of increasing interest in the field are those that lie within the radius valley (\cite{owen}, \cite{fulton}, \cite{Van_Eylen_2018}), i.e. between $1.5 - 2.0\,R_{\bigoplus}$,  whereby the dearth of planets in this region is theorized to be consistent with the intersection between super-Earths and sub-Neptunes or water-worlds. Planets smaller than $1.8\,R_{\oplus}$ could have thinner, possibly H$_2$-depleted atmospheres (e.g. \cite{Valencia_2010,laconte_15,Dressing2015}), whilst planets with radii larger than this threshold could possess volatile-rich atmospheres, H$_2$/He or H${_2}$O-rich in particular (e.g. \cite{Valencia_2013,19SeJaSt}). However, observational constraints, in the form of atmospheric transmission spectra, are needed to confirm these hypotheses and therefore the nature of these transitional planets. Furthermore, this population of small planets deviates from our understanding of planet atmospheres which has been mostly built upon our knowledge of hot-Jupiters and the solar system. For example, cross-sections that are utilized in the exoplanet field (based on the line lists provided by ExoMol \citep{Tennyson_exomol}), HITRAN \citep{gordon}, MoLLIST~\citep{MOLLIST} and HITEMP \citep{rothman}) are typically generated for atmospheres which are either dominated by H$_2$ and He, or air for Earth-like planets, in the case of HITRAN (see recent works, such as \cite{20ChRoRe.exo,20KiHeOr.exo,20PhTrBa.exo,jt819}).  Whilst this is appropriate for gas giant planets or Earth-like planets, such cross-sections do not include effects such as the self-broadening of heavy molecules like H$_2$O. When considering atmospheres heavier than H$_2$/He, such as is expected to be prevalent on super-Earths and water worlds \citep{19SeJaSt, Ducrot_2020, Crossfield_2020, Wunderlich_2020,19GhLi.broad}, this provides a non-negligible difference in the simulated (wavelength-dependent) atmospheric signal obtained during transit, knows as the \textit{transit depth}.

H$_2$ and He are light molecular species which only interact weakly with other molecules at long range and are therefore very inefficient
line broadeners. Water, conversely, is both heavier and possesses a significant dipole which leads to enhanced long-range 
interactions. Broadening by water vapour is known to be significant in the Earth's atmosphere even though it is generally only a trace
species. As transit spectroscopy probes regions of an exoplanet atmosphere as it approach optical thickness, the increased
line broadening by water can be expected to lead to significantly enhanced overall absorption compared to the case where only
H$_2$ and He are considered.

Previous works have investigated the effects of various choices when computing cross-sections for modelling exoplanet atmospheres, such as ~\cite{16HeMaxx,RocchettoPhD,20BaChGa.exo,21GhIyLi}. These investigations included the choice of broadening parameters, although these typically focus on H$_2$/He atmospheres and associated broadening parameters. Recently there has been some work more focused on cross-sections for heavier atmospheres, such as \cite{19GhLi.broad,22AnChEl}.

In this work we investigate the effects of including H$_2$O self-broadening, in addition to H$_2$/He-broadening, in the calculation of our water cross-sections which we use to model exoplanet atmospheres. We find that the transit depth is considerably affected; the largest difference being found for one of our case-study planets GJ 1214\,b, a $6.26$ $M_{\bigoplus}$, $2.85$ $R_{\bigoplus}$ super-Earth.

\section{Methodology}
\subsection{Transmission spectroscopy}

If the orbital plane of a planet around its host star is aligned approximately parallel to our line of sight with the system (analogous to $90^{\circ}$ inclination), the planet will transit in front of its star. Assuming that there is no atmosphere and that the planet is totally opaque to its incoming starlight, this transiting motion will cause a drop in the amount of stellar flux we receive from the host star. This change in detected light is known as the \emph{transit depth}, $\Delta F$, which is equal to the ratio between the surface area of the planet (as we view it, in 2D) and the surface area of the star. Since we assume both objects to be totally symmetrical, this reduces to:
        \begin{equation} \Delta F = \frac{F_{out} - F_{in}}{F_{out}} = \left(\frac{R_{p}}{R_{*}}\right)^{2} 
        \label{eq:tran_depth_opaque}
        \end{equation}
        which gives a measure of the relative change in flux as the planet blocks its starlight. This provides us with an observable quantity with which we can quantify the size of the planet, if we know the stellar properties, which can be derived from models.

        Now, if the planet possesses an atmosphere, an envelope of gas which surrounds the planet which is maintained by the planet's gravitational force, the molecules present will absorb, scatter and reflect incoming starlight, in addition to thermally emitting photons. Owing to the varied and distinct spectral characteristics of different molecules, how opaque a certain atmosphere is to incoming stellar flux will vary significantly with wavelength. This information is described by the quantity $\tau(\lambda)$, given in Eq. \ref{attenuated_intensity}, known as the \emph{optical depth}. Overall, regarding transmission spectroscopy, we can treat the atmosphere as a purely absorbing and single-scattering medium as a good (first-order) approximation for radiative transfer through the planetary atmosphere.

        Given an arbitrary path through the atmosphere for which radiation transmits with wavelength-dependent initial intensity $I_{\lambda}$, the transmitted radiance will be attenuated by absorption and scattering processes. We can denote this reduction in intensity as a function of path $ds$ as  $dI_{\lambda}/ds=-I_{\lambda}\sigma_{\lambda}\rho $, where $\sigma_{\lambda}$ is the total mass extinction cross section (the sum of the absorption and scattering cross sections) and $\rho$ is the density of the medium. Integrating up and using the fact that the optical depth as a function of atmospheric height is determined by summing the opacity contributions of all molecular species present, we recover the \emph{Beer-Bougert-Lambert Law}:
        \begin{equation}
            I_{\lambda}(z) = I_{\lambda}(0)e^{-\tau_{\lambda}(z)} \quad {\rm with} \quad \tau_{\lambda}(z) = \sum_{m}\int_{z}^{z_{\infty}} \sigma_{m, \lambda}(z')\chi_{m}(z')\rho(z')dz',
            \label{attenuated_intensity}
        \end{equation} 
        where $\chi_{m}$ and $\rho$ are the column density of a given molecular species and the number density of the atmosphere, respectively.

        We may now rewrite Eq.~(\ref{eq:tran_depth_opaque}) as:
        \begin{equation}
            \Delta F = \frac{F_{out} - F_{in}}{F_{out}} = \left(\frac{R_{p}+h_{\lambda}}{R_{*}}\right)^{2} \approx \frac{R_{p}^2+2R_{p}h_{\lambda}}{R_{*}^2} \qquad O(h_{\lambda})
            \label{eq:atmos_transit_depth}
        \end{equation}
        where we may describe the atmospheric height function as:
        \begin{equation}
            2R_{p}h_{z} = 2\int_{0}^{z_{max}} (R_{p}+z)(1-e^{-\tau_{\lambda}(z)}) dz,
            \label{eq:atmost_height_function}
        \end{equation} 
        where $z_{max}$ denotes the height of the atmosphere.

Using this formalism, the transit depth of an atmosphere-bearing planet for any given wavelength may be calculated, using derived \textit{cross-sections} (temperature, pressure and wavelength dependent) for a given molecular species. If we populate a model atmosphere with a given temperature profile, pressure profile, chemical species abundances and a specified cloud distribution we may generate a transmission spectrum ($\Delta F$ vs. $\lambda$) for the atmosphere, thereby \textit{forward-modelling} it. To date, there is extensive literature pertaining to the collection of transmission data, alongside analysis of the generated transmission spectra, for a wide variety of exoplanets; from hot-Jupiters \cite{Skaf_2020,Pinhas_ten_HJ_clouds,Sing_2015_popstudy} to habitable-zone super-Earths \cite{de_Wit_2018, Tsiaras_2019_k2-18, Edwards_2020}.  

\subsection{Broadening parameters}

 The Voigt profile is commonly used to represent line broadening in exoplanet atmospheres, which is a convolution of the temperature-dependent Gaussian line profile and the pressure-dependent (and therefore dependent on broadening species) Lorentzian profile. The equation for the Lorentzian line width (HWHM) for a given pressure $P$ and temperature $T$, is given by: 
\begin{equation}\label{eq:gamma_L}
\gamma_L = \gamma \left( \frac{T_0}{T}  \right)^{n} \frac{P}{P_0}.
\end{equation}
Here, $T_0$ and $P_0$ are the reference temperature and pressure, whilst $\gamma$ and $n$ are the reference HWHM and temperature exponent, respectively. The latter two terms are known as pressure-broadening parameters and are dependent primarily on the molecular species being broadened and the species inducing it. Therefore these are the parameters which, upon variation, enable us to study the self-broadening effects of water considered in this study; considering atmospheres comprising varying levels of H$_2$O with respect to H$_2$ and He. We note that the power law for temperature dependence which we assume in this work may not work well over large temperature ranges, including those temperatures of the atmospheres we are modeling. This has been demonstrated by works such as \cite{18GaVi}, who developed the more advanced  Gamache–Vispoel double power law (DPL) model. Others, such as Wcis{\l}o {\it et al.} \cite{16WcGoTr.H2} highlight non-Voigt effects to line broadening. We will consider updating our cross-sections using these more robust line-shape models in the future. We also expect to have more robust values for the self-broadening parameters of H$_2$O and other molecular species, as a result of ongoing work from projects such as ExoMol~\citep{20TeYuAl} and HITRAN~\citep{19TaKoRo.broad} and others~\citep{18HaTrAr,20StThCy.broad}, which we would like to incorporate into our cross-sections in the future. For the present study we use the Voigt profile of Eq.~\ref{eq:gamma_L}.
The self-broadened H$_2$O cross-sections used in this study were computed using ExoCross \citep{ExoCross}, as were the H$_2$/He-broadened cross-sections for H$_2$O. The latter are similar to those from the ExoMolOP database~\citep{20ChRoRe.exo} but with $J$-dependent broadening parameters (where $J$ is the rotational angular momentum quantum number) from the ExoMol website\footnote{\url{https://www.exomol.com/data/data-types/broadening_coefficients/H2O/}} used for H$_2$ and He broadening \cite{jt669,jt684}. All H$_2$O cross-sections presented here use the ExoMol POKAZATEL line list~\citep{polyansky_h2o}. We computed the line wings out to 500 Voigt widths in all cases, out to a maximum of 25~cm$^{-1}$. The values of $\gamma$ and $n$ for the self-broadening of H$_2$O used in the present study are detailed below.

\subsubsection{Self-broadened half-width, $\gamma_{\rm H_2O}$}

There are many literature sources with broadening parameters for self-broadening of H$_2$O. For example, Gamache and Hartmann \cite{04GaHa} compiled and compared various parameters related to H$_2$O line shape, including values for the half-width $\gamma$ for self-H$_2$O broadening. There are over 47,000 lines in their database, with values of $\gamma_{\rm H_2O}$ ranging from 0.108~-~0.805~cm$^{-1}$~atm$^{-1}$. The 2020 release of the HITRAN database~\citep{HITRAN2020}, who follow a ``diet'' procedure \cite{07GoRoGa}, provide an update of this 2004 broadening measurement database of \cite{04GaHa}.  A simple average of all values of the main isotopologue of H$_2$O (with no weighting) from HITRAN2020 for $\gamma_{\rm H_2O}$ yields a value of 0.35~cm$^{-1}$~atm$^{-1}$. There are many other works with available broadening values: for example, \cite{05Toth} and \cite{17LoBiWa} both present a number of values for $\gamma_{\rm H_2O}$ for a few thousand lines each. They report average values of $\gamma_{\rm H_2O}$~=~0.4 and between 0.1 and 0.5~cm$^{-1}$~atm$^{-1}$, respectively.

As noted above, for the present study we extract the broadening parameters from the HITRAN2020~\citep{HITRAN2020} database as a function of rotational angular momentum quantum number $J$, computing an average value of $\gamma$ for each value of $J$. The data extends up to a maximum of $J$=26 and varies between 0.1~cm$^{-1}$~atm$^{-1}$ for high $J$ to 0.5~cm$^{-1}$~atm$^{-1}$ for low $J$.

\subsubsection{Temperature exponent $n_{H_2O}$ for self-broadened half-width}

HITRAN2020~\citep{HITRAN2020} currently only include $\gamma_{\rm self}$ and not $n_{\rm self}$ values for H$_2$O, due to the large effort required to validate and populate such parameters into their database. For now we use averaged $J$-dependent values for $\gamma_{\rm self}$ from HITRAN2020~\citep{HITRAN2020} and an apparently typical value of  $n_{\rm self}$=0.7 for H$_2$O.
The focus of some ongoing and future work is to update the ExoMol~\citep{20TeYuAl} and ExoMolOP~\citep{20ChRoRe.exo} databases to include broadening parameters in a more comprehensive way. This is not a simple undertaking: the H$_2$O line list used in this work, for example, ExoMol POKAZATEL~\citep{polyansky_h2o}, contains 6~billion transitions between 800,000 energy levels, with even larger line lists required to describe larger species, see Table 13 of \cite{20ChRoRe.exo}.

Although it is beyond the scope of the current study to perform a comprehensive assessment of available temperature exponents for self-broadening of H$_2$O, there are a number of studies in the literature which have focused on analysing water vapor spectra at various temperatures in order to determine the temperature dependence $n$ of the self-broadened half-width $\gamma$. Here, we summarize the results from a selection of works, but note that this is not a comprehensive sample. 
Grossmann and Browell \cite{89GrBr} analyzed spectra of water vapor in the 720~nm region, finding an average value of $n$=0.75. Studies such as \cite{94Markov,87BaGoKh,89BaGoKh} all analyzed particular rotational lines in the low-wavenumber region of the spectra, between around 250~-~390~K. They find values of $n_{\rm self}$ of 0.62, 0.89 and 0.85, respectively. Both \cite{87BaGoKh,89BaGoKh} look at various broadeners, including self.  They find that both $\gamma$ and $n$ are generally larger for the cases where H$_2$O is the broadener, in comparison to N$_2$, O$_2$ or Ar as broadeners ($n$ is 0.52, 0.64, 0.49 for those cases, respectively for example in \cite{87BaGoKh}). 
Alder-Golden {\it et al.} \cite{92AdLeGo} analyzed low-$J$ lines of H$_2$O close to 12,200~cm$^{-1}$ in the 330~-~540~K temperature range. They find an average value of $\gamma_{\rm self}$ at 296~K of 0.456~cm$^{-1}$~atm$^{-1}$ (compared to 0.095~cm$^{-1}$~atm$^{-1}$ for air-broadening). The temperature coefficient $n$ for self-broadening of H$_2$O was found to be 0.9 on average across the spectral region measured. Podobedov {\it et al.}
\cite{04PoPlFr} analyzed several lines of H$_2$O in the region around 12~-~52~cm~$^{-1}$ for temperatures between 263~-~340~K and over a pressure range from 0.0003~-~0.014~bar. The $J$- and $T$-dependent values of $\gamma$ were found to vary between 0.67~-~1.07~cm$^{-1}$~atm$^{-1}$. They found $J$-dependent values of $n_{\rm self}$ between 0.56 and 0.81.

Table~\ref{t:HITRAN2020_broad} gives some average values extracted from the HITRAN 2020 \citep{HITRAN2020} database for various molecules and broadeners. The H$_2$- and He-broadened average values used in the ExoMolOP database \citep{20ChRoRe.exo} (a database tailored for modelling ``hot Jupiter''-type exoplanet atmospheres) are also included for reference. It can be seen that in general that the largest $\gamma$ and $n$ values occur when H$_2$O is the broadener, which for $\gamma_{\rm H_2O}$ is an order or magnitude larger than H$_2$- and He-broadening. The contrast between H$_2$- or He-broadening and H$_2$O-broadening, however, does not appear to be so large for other species as for water vapor. A similar observation was noted by Gharib-Nezhad and Line \cite{19GhLi.broad}, who highlight that self-broadening for H$_2$O is typically up to 7 times larger than H$_2$/He broadening based on their compilation of literature values.

\begin{figure}
    \centering
    \includegraphics[width=0.45\textwidth]{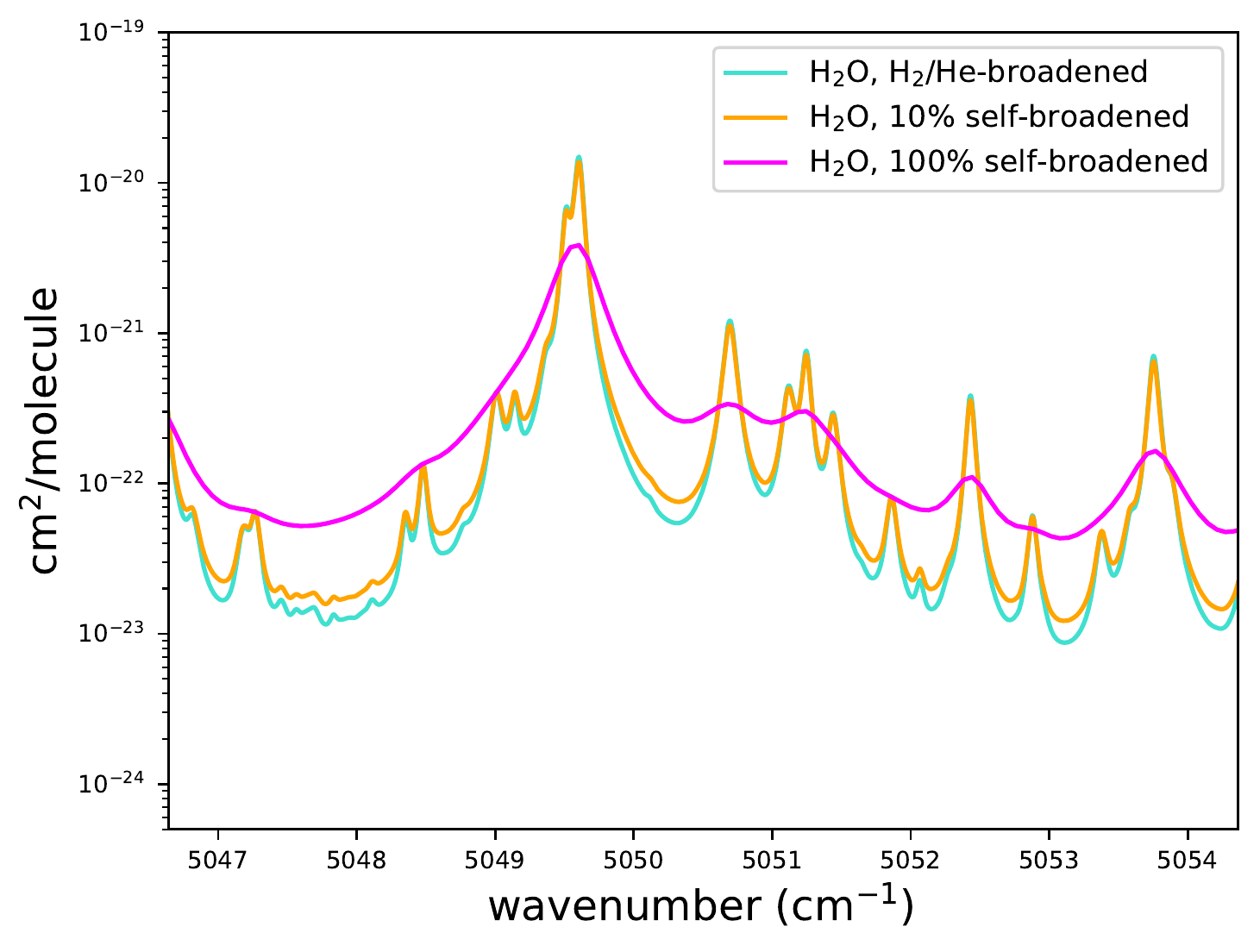}
       \includegraphics[width=0.45\textwidth]{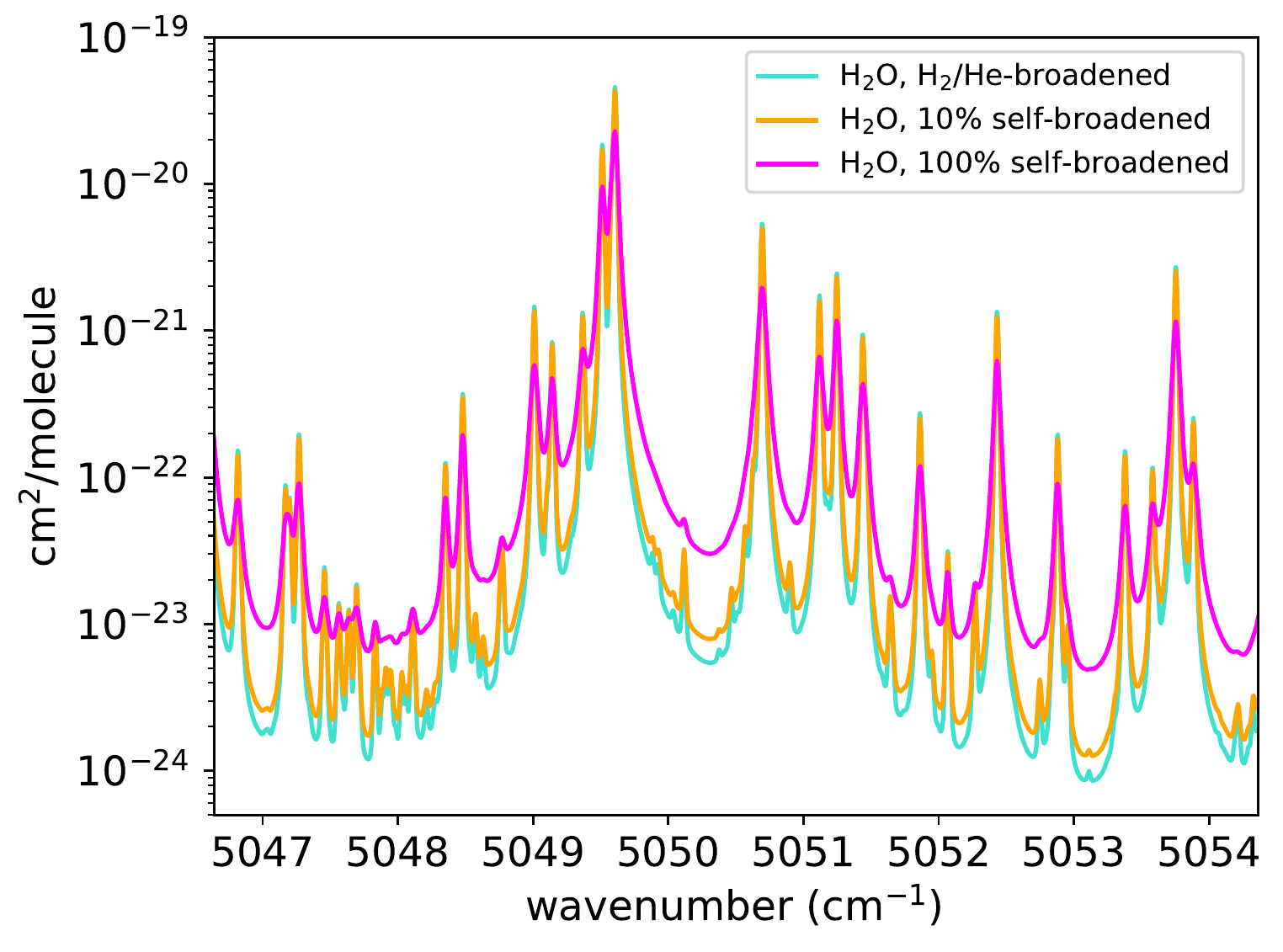}
    \caption{Cross-sections computed using the self-broadening parameters for H$_2$O, for the abundances of water vapor given in the legend. We compare the self-broadened cross-sections (orange, pink) with the H$_2$/He-only broadened cross-sections (blue). Left: atmospheric pressure of $1$ bar and right: atmospheric pressure of $10^{-1}$ bar. In an atmosphere with 10\% H$_2$O, we assume the remaining 90\% atmosphere is comprized of H$_2$/He in solar abundances and that the broadening is therefore 90\% dominated by H$_2$/He and the remaining 10\% from H$_2$O self-broadening. In this case we create a combined cross-section which includes 90\% of the self-broadened and 10\% of the H2/He-broadened cross-section for a given pressure and temperature. }
    \label{fig:xsecs_pressures}
\end{figure}

 In Figure \ref{fig:xsecs_pressures} we present our derived H$_2$O-broadened cross-sections at $T = 600$ K for various pressures, in comparison to the standard H$_2$/He-broadened ones; in both cases it is clear that H$_2$O self-broadening widens the profiles of each absorption peak.

\subsection{Simulating transmission spectra}

\begin{figure}
    \centering
    \includegraphics{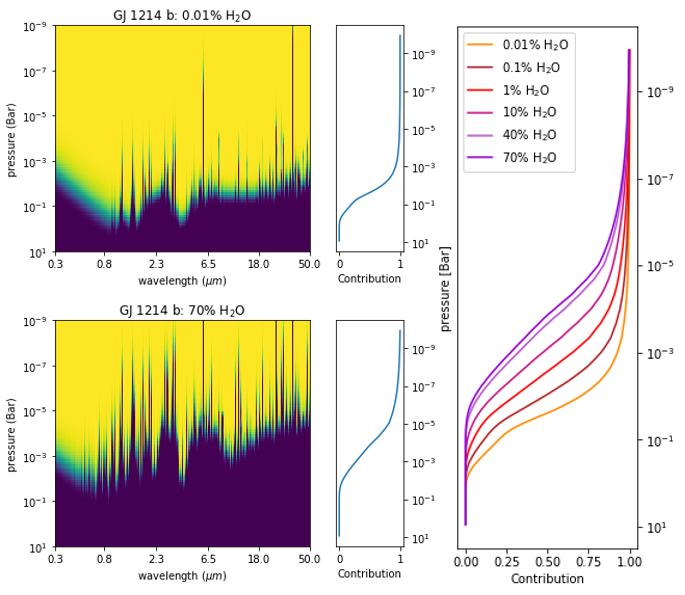}
    \caption{The contribution function, $\frac{d\tau}{dP}$, is defined as the wavelength-averaged variation in the optical depth, $\tau$ with pressure, $P$. On the LHS, $\tau$ is plotted in $\mu - P$ space; for which the contribution $\frac{d\tau}{dP}$ is normalized and displayed in the central panel; whilst on the RHS this contribution is plotted for varying water vapor abundance; illustrating that deeper pressures can be probed in lighter atmospheres. }
    \label{fig:contributionFunctions}
\end{figure}

Eq.~(\ref{eq:gamma_L}) exhibits a linear relationship for $\gamma_{L}$ with atmospheric pressure, $P.$ Hence, we anticipate the strongest broadening effect to contribute deeper in the atmospheres we wish to model. Consequently, the abundance of water vapor in the atmosphere should affect our results, since although a decrease in molecule number density ought to minimize the overall opacity contribution induced by self-broadening, lighter atmospheres with low-mean-molecular weight allow us to probe deeper pressures; this is illustrated in Figure \ref{fig:contributionFunctions}, where the contribution function, $\frac{d\tau}{dP}$, is defined as the wavelength-averaged variation in the optical depth, $\tau$ with pressure, $P$. This informs our implementation of atmospheric models endowed with a variety of water vapor abundances, in order to examine for which abundances the self-broadening effects of H$_2$O are both most prominent and most observable with future space missions.

In order to simulate forward models of transmission spectra for GJ 1132\,b and GJ 1214\,b, this analysis was performed using the publicly available retrieval suite TauREx 3.0 \citep{waldmann_2,waldmann_1,al-refaie_taurex3}. For the stellar parameters and the planet mass, we used the values from \cite{Bonfils_2018} and \cite{Harpsoe_GJ1214} as given in Table \ref{tab:params} a). In our runs we assumed that the planets possess a range of different water-based atmospheres, with fill gas abundance ratio given by $V_{{\rm H}_2{\rm O}}$ = $x$, with the chosen values of $x$ given in Table \ref{tab:params} b) and the remainder of the molecular abundance made up of H$_2$ and He, as specified. 
Additionally, we included the collision induced absorption (CIA) from H$_2$-H$_2$ \citep{abel_h2-h2, fletcher_h2-h2} and H$_2$-He \citep{abel_h2-he}, as well as Rayleigh scattering for all molecules. Finally, our simulated atmospheres are cloud-free, isothermal and have molecular abundance profiles which are constant with altitude. The assumptions of isothermal and iso-chemical atmospheres hold for interpreting current data (\cite{Edwards_2020, anisman2020}). While these approximations might be too simplistic to interpret accurately JWST and Ariel data (\cite{Changeat_2019, rocchetto}), they will not change the conclusions of our paper. 
For each of the two planets, two types of spectra were generated: one using the standard H$_2$/He-broadened cross-sections only and another with the H$_2$O self-broadened cross-sections included for the various percentages of H$_2$O modeled in the atmosphere, as described in Section 2.2 and illustrated in Figure~\ref{fig:xsecs_pressures}.

\begin{table}
\caption{a) stellar and planetary parameters for two small planets, for input into TauREx 3.0, derived from \cite{Bonfils_2018}, \cite{Harpsoe_GJ1214}, b) list of the forward-modeled parameters, their values and the scaling used.}
\centering
\begin{tabular}{cccc}
\hline\hline
\multicolumn{4}{c}{a) stellar \& planetary parameters}\\ 
\hline
    parameter  & GJ 1132\,b & GJ 1214\,b &\\ \hline
    $T_{*}$ [K]   & 3270 & 3026&\\
    $R_{*}$ [$R_\odot$] &  0.207 & 0.220&\\
    $M_{*}$ [$M_\odot$] & 0.181 & 0.176&\\
    $M_{p}$ [$M_{\bigoplus}$] & 1.66 & 6.26 &\\
    $R_{p}$ [$R_{\bigoplus}$] & 1.13 & 2.85&\\
    $P_{orbital}$ [days] & 1.63& 1.58&\\
    \end{tabular}
    \begin{tabular}{cccc}
  \hline \hline
\multicolumn{4}{c}{b) forward model parameters}\\ \hline
\mbox{parameter}        & \mbox{GJ 1132\,b}   & \mbox{GJ 1214\,b} & \mbox{type} \\
\hline
\mbox{$P_{clouds}$}    & \mbox{None}  & \mbox{None} & \mbox{opaque} \\
\mbox{$T$ [K]}        & \mbox{$500$} & \mbox{$600$}  & \mbox{iso.}\\
\mbox{$\log_{10}{V_{{\rm H}_2{\rm O}}}$} $\in$& [-2, 2]  & [-2, 2]    & \mbox{fill}\\
\mbox{$V_{\rm He}$}& \mbox{$1e-7$}  & \mbox{$1e-7$}  & \mbox{trace} \\
\hline \hline
 \end{tabular}
\label{tab:params}
\end{table}

\section{Results}

\subsection{Forward-modelling of transmission spectra}

\begin{figure*}[h]
\centering
\includegraphics[width=0.32\textwidth]{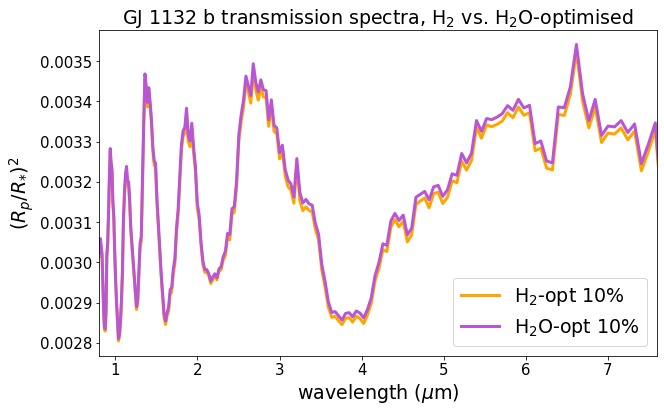}
\includegraphics[width=0.32\textwidth]{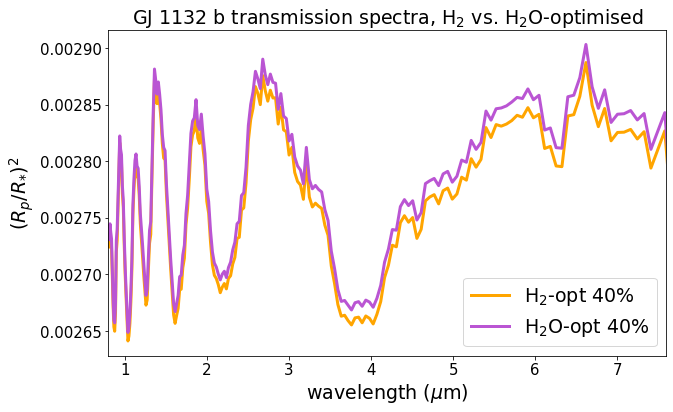}
\includegraphics[width=0.32\textwidth]{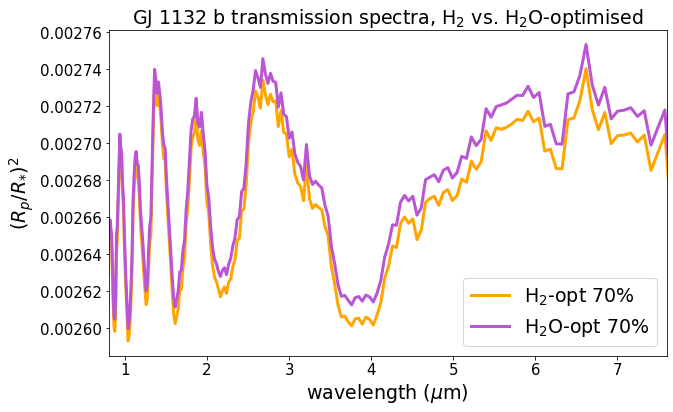}
\includegraphics[width=0.31\textwidth]{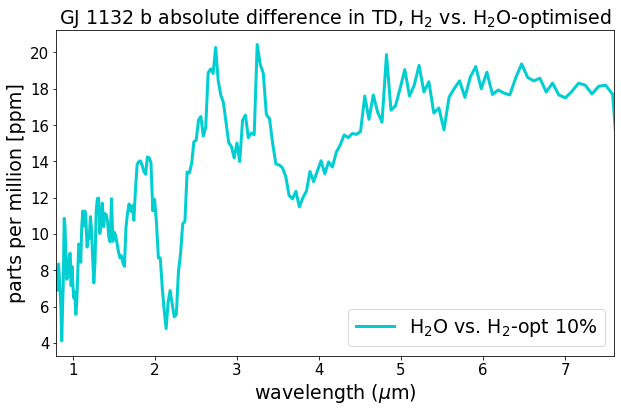}
\includegraphics[width=0.31\textwidth]{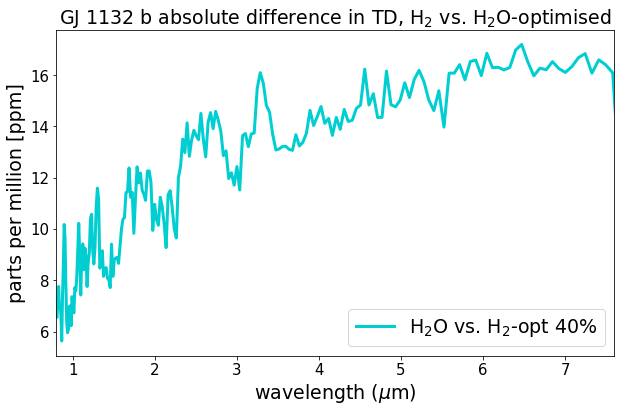}
\includegraphics[width=0.31\textwidth]{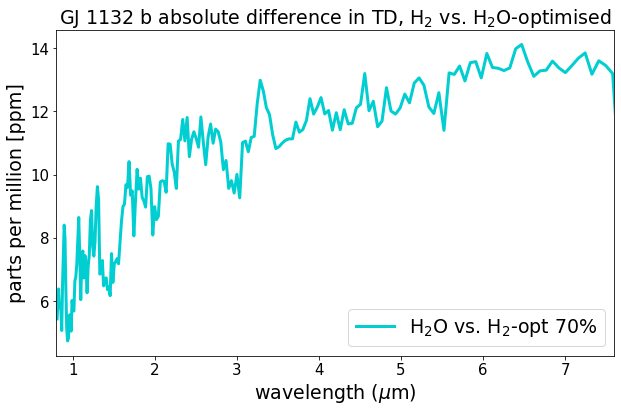}

\includegraphics[width=0.32\textwidth]{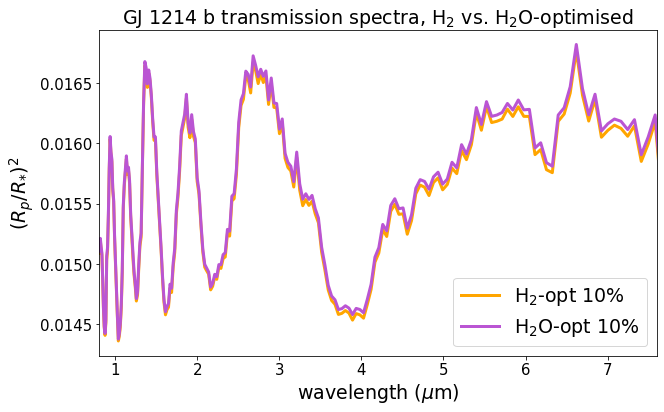}
\includegraphics[width=0.32\textwidth]{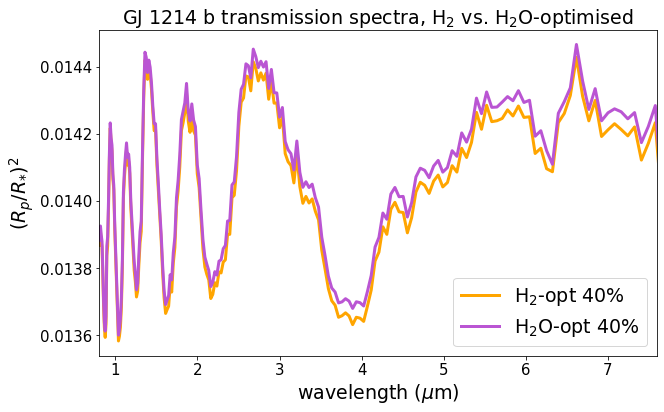}
\includegraphics[width=0.32\textwidth]{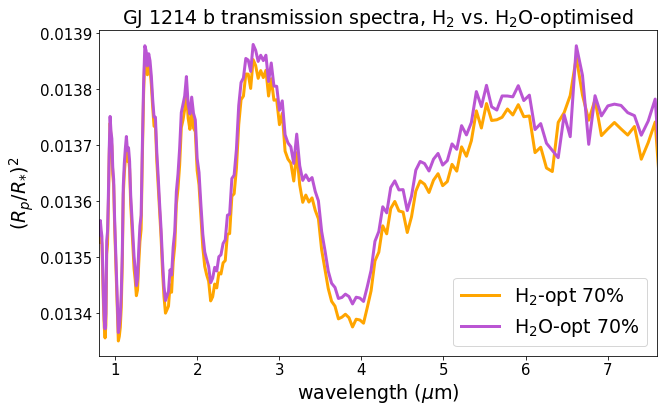}
\includegraphics[width=0.31\textwidth]{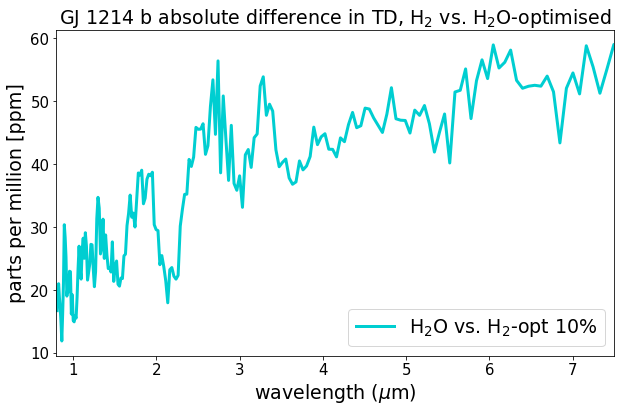}
\includegraphics[width=0.31\textwidth]{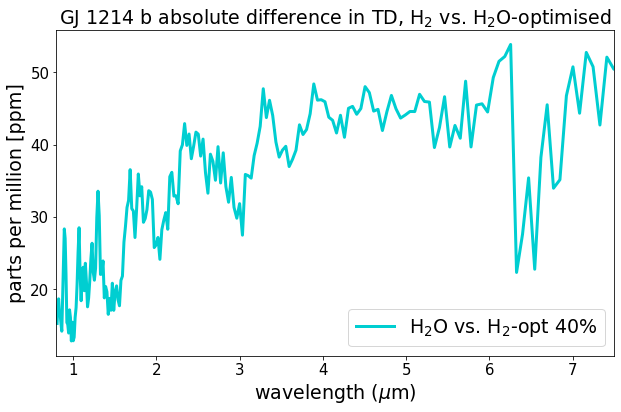}
\includegraphics[width=0.31\textwidth]{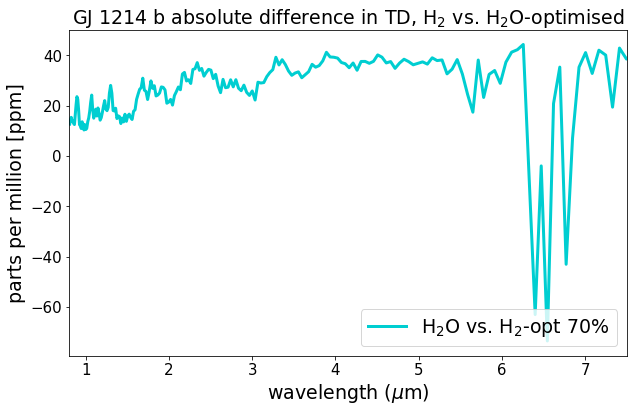}

\caption{Individual over-plots of the (resolution 200) spectra produced with both H$_2$ and H$_2$O-broadened cross-sections for each planet, for varying water abundance, with absolute differences beneath. Top: GJ 1132\,b (500 K), bottom: GJ 1214\,b (600 K).}
\label{fig:crossections_overplots}
\end{figure*}

We generate transmission spectra at a native resolution of 15,000 for both planets, before binning down to a nominal resolution and signal-to-noise ratio of 200 and 10, respectively, with TauREx 3.0 (enabling eventual comparison with Ariel and JWST errorbars). We utilize both the H$_2$/He- and the H$_2$O-broadened cross-sections described in Section 2.1, yielding the spectra given in Figure \ref{fig:crossections_overplots}, with  planetary and stellar parameters described in detail in Table \ref{tab:params}. In comparison with spectra produced using H$_2$/He-broadened cross-sections, it is evident that by using cross-sections calculated for the water-dominated atmospheres the spectral features are amplified due to the additional absorption achieved by including self-broadening of H$_2$O. In order to quantify these wavelength-dependent differences, these sets of spectra which are over-plotted the top panels of Figure \ref{fig:crossections_overplots} have been subtracted to obtain the results presented in the lower panels, from which we determine maximum absolute differences in transit depth for the lightest secondary-type atmospheres (10\% H$_2$O); of 60 and 20 ppm for GJ 1214\,b and GJ 1132\,b, respectively. 

\subsection{Ariel \& JWST error-bars}
During its primary mission, Ariel will survey the atmospheres of 1000 exoplanets \citep{tinetti_ariel,tinetti_ariel2} and many of these targets could be in the Super-Earth and Sub-Neptune regime \citep{edwards_ariel}. Meanwhile, during the Guaranteed Time Observations (GTO) and first cycle of the General Observing (GO), around 70 planets will be observed with JWST \citep{gardner_jwst}. Nearly half of these JWST targets have a radius of $\leq 2.5R_\oplus$ and, therefore, may not have a H$_2$-dominated atmosphere. During its lifetime, JWST is expected to observe a couple of hundred exoplanets \citep[e.g.][]{cowan}.

To investigate the detectability of broadening-induced differences with future instruments, we generated error bars for the simulated spectra. We then compared the size of these uncertainties to the absolute differences between the H$_2$- and H$_2$O-broadened spectra. For Ariel, we generated error bars using ArielRad \citep{mugnai} while, for JWST, we used a modified version of the radiometric tool described in \citet{edwards_terminus} which utilizes the JWST instrument parameters from Pandeia \citep{pontoppidan_pandeia}.

\begin{figure}
\centering
\includegraphics[width=0.75\textwidth]{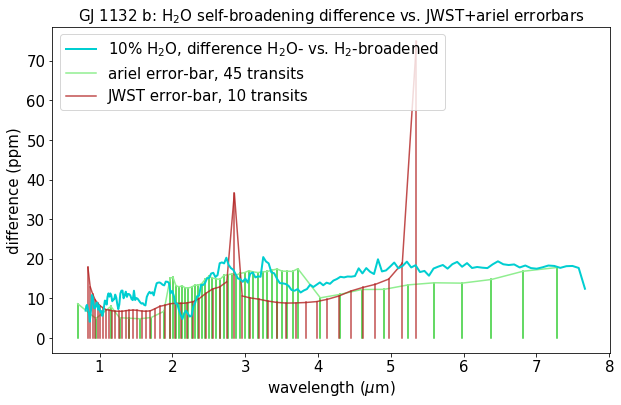}
\includegraphics[width=0.75\textwidth]{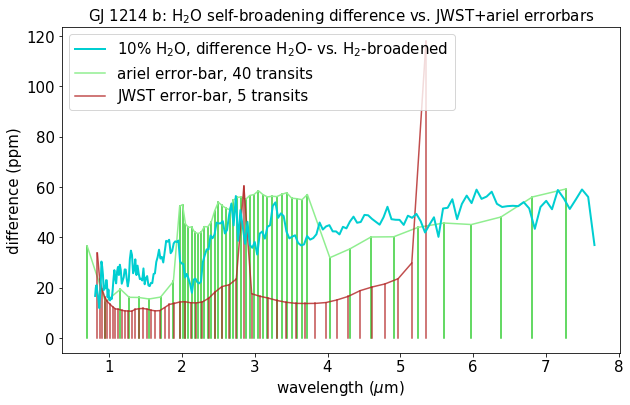}
\caption{Comparison of simulated error bars for observation of a) GJ 1132\,b and b) GJ 1214\,b with future-missions Ariel and JWST with the absolute wavelength-dependent differences in transit depth on the simulated transmission spectra, using H$_2$- vs. H$_2$O-broadened cross-sections. In both cases these broadening-induced differences should be directly observable in the near-future.}
\label{fig:errorBars}
\end{figure}

For JWST we modeled observations with NIRISS GR700XD (0.8 - 2.8 $\mu m$) and NIRSpec G395M (2.9 - 5.3 $\mu m$), whilst for Ariel, which provides simultaneous coverage from 0.5 - 7.8 $\mu m$, we simulated error bars at tier 2 resolution. Presented in Figure \ref{fig:errorBars}, for GJ\,1132\,b and GJ\,1214\,b, we observe that both instruments will be sensitive enough to reveal such differences when integrating multiple transits, specifically  45 and 40 in the case of Ariel (tier 2) and 10 and 5 in the case of JWST for GJ 1132\,b and GJ 1214\,b, respectively. We note that both these planets will be studied by JWST in the first cycle of observations.

\section{Discussion}

In all spectra, an increase in $\mu$, corresponding to increased volume mixing ratio of water, corresponds to reduced scale height, since $H=\frac{kT}{\mu g},$ (where $\mu$ is the mean molecular weight of the atmosphere, $g$ is the planet gravity and $k$ is the Boltzmann constant). For the individual atmospheres we observe a saturation of features as the atmospheric mean-molecular-weight gets larger. In tandem, an increase in mean-molecular weight of the atmosphere results in a decrease in the atmospheric pressure at which we can probe, as illustrated in Figure \ref{fig:contributionFunctions}. Thus, the largest observed differences in transit depth vs. wavelength are found in the lightest secondary-type atmospheres, namely those with 10\% water vapor, for all employed temperatures. As for the two planets considered, the largest transit depth increases are observed in GJ 1214\,b, with an equilibrium temperature of 600 K, as opposed to HAT-P-11\,b, which possesses an equilibrium temperature of 900 K.

Although the Lorentzian line profile for H$_2$O self-broadening is approximately inversely proportional to the atmospheric temperature, low-temperature planets have smaller global transit depths, due to the reduction in scale height the temperature induces. Hence, there is a trade-off between the strength of the broadening effect and detectability of such effects. Our work finds that the  atmospheric temperatures for such differences to become detectable with future missions, as discussed, is thus $\sim 600$ K.
\\

Thus, consideration and utilization of self-broadening effects induced by not just H$_2$O, but all large-$\mu$ molecules should have a stronger impact on the medium-cool atmospheres of temperate planets, as well as those with low-mean-molecular-weight secondary atmospheres, namely those closer to the threshold abundance of around 10\% non-H$_2$ or -He which we define to the boundary between primary and secondary type atmospheres; precisely those planets which sit in the transition region between super-Earths and sub-Neptunes, whose atmospheres remain illusive and out-of-reach with current telescopes, due to known limits on both the signal-to-noise and resolution (\cite{Edwards_2020}. The next generation of telescopes will widen and deepen our spectral view, both in terms of signal-to-noise and resolution, but also in wavelength coverage; hence these problems should be easily tackled with Ariel and JWST. It is therefore especially imperative for the accuracy of cross-section data to compete with the level of precision obtainable with these future facilities.
\\

Standard H$_2$/He-broadened cross-sections which are in widespread use by the exoplanet community are simply non-optimal for the study of secondary-type, heavy atmospheres dominated by non-H$_2$/He species due to the fact that these opacities are calculated with respect to a nominal atmosphere which is dominated by H$_2$ and He. At present, due to the computationally-intensive nature of numerically evaluating absorption strengths for molecular transitions and interactions for a given atmospheric species, specifying a grid of temperature and pressure values is necessary to obtain computable opacities for input into forward models and atmospheric retrieval algorithms. It is well known that extrapolating opacities above the temperature or pressure grid on which they were computed can be problematic (e.g. \cite{Tinetti_2012}, \cite{Yurchenko_2014}), largely due to the dominance of transitions that originate in energy levels above the ground-state in higher temperature regimes. In this work we demonstrate that including self-broadening effects into the calculation of the absorption cross-section of water markedly affects the simulated transit depth of small-planet exoplanet atmospheres. Moreover, we are the first in the exoplanet field to illustrate that these generated differences are detectable with the near-future space missions JWST \& Ariel, by explicitly simulating spectra with error bars for these instruments. We prove that self-broadening is necessary to account for in these calculations. As a community, our long-term goal should be to develop cross-section functions,  explicitly derived for pressure-temperature grids and as a function of molecular abundances, with all intra- and inter-molecular effects, like self-broadening, included. 

\section{Conclusion}
In summary, it is evident that accounting for previously thought-to-be negligible absorption contributions, such as the self-broadening exhibited by H$_2$O in our opacity functions, will alter simulated transit depths by as much as 60 ppm. These differences sit above the noise level for a reasonable number of transit observations with the near-future space telescopes JWST and Ariel for the two small planets considered: GJ 1132\,b and GJ 1214\,b. Our quantification of the transit depth differences found by producing and utilising cross-sections which include the absorption contribution induced by H$_2$O self-broadening motivate further progress in not only refining such broadening parameters, but also developing opacities for a variety of molecular species expected to be found in the atmospheres of small planets, which may also be dominated by more than one heavy molecule. This is not an easy undertaking, due to the vast amount of work done and ongoing in the field of line shapes (including half widths $\gamma_{\rm H_2O}$ and temperature dependence $n_{\rm H_2O}$; see, for example, \cite{04GaHa, 19TaKoRo.broad,HITRAN2020,05Toth} and references therein). It is paramount that the field of exoplanet spectroscopy moves towards the use of more adaptive cross-sections, built as functions not only of temperature and pressure but also of molecular abundance, as we have illustrated specifically for the case of H$_2$O.

\section{Acknowledgements}
\textit{Funding:} We acknowledge funding from the ERC through Consolidator grant ExoAI (GA 758892) and Advanced Grant
ExoMolHD (GA 883830), as well as from STFC grants ST/P000282/1, ST/P002153/1, ST/S002634/1 and ST/T001836/1. BE is a Laureate of the Paris Region fellowship programme which is supported by the Ile-de-France Region and has received funding under the Horizon 2020 innovation framework programme and the Marie Sklodowska-Curie grant agreement no. 945298.

\bibliographystyle{elsarticle-num-names} 
\bibliography{elsarticle-template-num}

\begin{table*}[h]
	\caption{Average values of $\gamma$ and $n$ for various species where parameters are available from HITRAN-2020~\citep{HITRAN2020}. The custom data search from HITRANonline as described in \cite{16HiGoKo} was used in order to extract the relevant parameters. Where ExoMolOP~\citep{20ChRoRe.exo} is labelled as the source (where only H$_2$ and He-broadening was considered), the citations used to find the averaged values are listed in the footnote to this table.}
	\label{t:HITRAN2020_broad} 
	\centering  
	\begin{tabular}{lllll}
		\hline\hline
		\hline
		\rule{0pt}{3ex}Species& Broadener & $\gamma$ & $n$ & Source \\
		\hline\hline
\rule{0pt}{3ex}CO$_2$	&	H$_2$O	&	0.14	&	0.79	&	HITRAN	\\
&	Self	&	0.09	&	0.64	&	HITRAN\\		
&	Air	&	0.07	&	0.71	&	HITRAN\\		
&	H$_2$	&	0.11	&	0.58&	HITRAN	\\		
&	He	&0.06	&	0.3	&	HITRAN	\\		
&	H$_2$	&	0.09	&	0.59&	ExoMolOP	\\		
&	He	&0.04	&	0.44	&	ExoMolOP	\\		
\hline									
\rule{0pt}{3ex}CO	&	H$_2$O	&	0.09	&	0.68	&	HITRAN	\\
&	Self	&	0.06	&	-	&	HITRAN	\\	
&	Air	&	0.05	&	0.7	&	HITRAN\\		
&	CO$_2$	&	0.06	&	0.66	&	HITRAN\\		
&	H$_2$	&	0.07	&	0.58	&	HITRAN	\\	
&	He	&	0.05	&	0.54	&	HITRAN	\\	
&	H$_2$	&	0.07	&	0.65	&	ExoMolOP	\\	
&	He	&	0.05	&	0.6	&	ExoMolOP	\\	
\hline									
\rule{0pt}{3ex}CH$_4$	&	H$_2$O	&	0.07	&	0.85	&	HITRAN	\\
&	Self	&	0.07	&	-	&	HITRAN	\\	
&	Air	&	0.05	&	0.67	&	HITRAN\\		
&	H$_2$	&	0.06	&	0.6	&	ExoMolOP	\\	
&	He	&	0.03	&	0.3	&	ExoMolOP	\\	
\hline									
\rule{0pt}{3ex}H$_2$O	&	Self	&	0.35	&	-	&	HITRAN	\\
&	Air	&	0.07	&	0.62	&	HITRAN\\		
&	H$_2$	&	0.06	&	0.2	&	ExoMolOP	\\	
&	He	&	0.01	&	0.13	&	ExoMolOP	\\	
				\hline\hline
\end{tabular}
\mbox{}\\

{\flushleft
CH$_4$:  \cite{90VaChxx.CH4,92Pine.CH4,98FoJeSt.CH4,72VaTexx.CH4,89VaChxx.CH4,01GrFiTo.CH4,04GaGrGr.CH4,17MaBuWe,19GhHeBe.ch4} \\
CO:  \cite{jt544,15LiGoRo.CO,18MuSuAl.co,05MaDeMa.co,16PrEsRo.co,98SiDuBe.co} \\
H$_2$O: \cite{jt483,12VoLaLu.H2O,jt669,08SoStxx.H2O,09SoStxx.H2O,13PeSoSo.H2O,12PeSoSt.H2O,16PeSoSo.H2O,19GaViRe.h2o} \\
CO$_2$ assumed parameters of C$_2$H$_2$: \cite{16WiGoKo.pb} \\
}
\end{table*}

\end{document}